\UseRawInputEncoding
\documentclass[a4paper,11pt]{article}
\pdfoutput=1 

\usepackage{jcappub} 

\usepackage[T1]{fontenc} 

\usepackage{graphicx}	
\usepackage{amsmath}	
\usepackage{amssymb}	
\usepackage{mathtools}
\usepackage{mathabx}
\DeclarePairedDelimiter\floor{\lfloor}{\rfloor}
\usepackage{bm} 

\title{\boldmath Super-resolving Dark Matter Halos using Generative Deep Learning}

\author[a,1]{D. Schaurecker,\note{Corresponding author.}}
\author[b,c]{Y. Li,}
\author[d]{J. Tinker}
\author[b,d,e,f]{S. Ho}
\author[a]{and A. Refregier}

\affiliation[a]{Institute for Particle Physics and Astrophysics, ETH Zurich, Switzerland}
\affiliation[b]{Center for Computational Astrophysics, Flatiron Institute - Simons Foundation, New York City, USA}
\affiliation[c]{Center for Computational Mathematics, Flatiron Institute - Simons Foundation, New York City, USA}
\affiliation[d]{Center for Cosmology and Particle Physics, New York University, New York City, USA}
\affiliation[e]{Physics Department, Carnegie Mellon University, Pittsburgh, USA}
\affiliation[f]{Department of Physics, Princeton University, Princeton, USA}

\emailAdd{dschaurecker@gmail.com}
\emailAdd{eelregit@gmail.com}
\emailAdd{jlt12@nyu.edu}
\emailAdd{shirleyho@flatironinstitute.org}
\emailAdd{alexandre.refregier@phys.ethz.ch}

\abstract{Generative deep learning methods built upon Convolutional Neural Networks (CNNs) provide great tools for predicting non-linear structure in cosmology. In this work we predict high resolution dark matter halos from large scale, low resolution dark matter only simulations. This is achieved by mapping lower resolution to higher resolution density fields of simulations sharing the same cosmology, initial conditions and box-sizes. To resolve structure down to a factor of 8 increase in mass resolution, we use a variation of U-Net with a conditional Generative Adversarial Network (GAN), generating output that visually and statistically matches the high resolution target extremely well. This suggests that our method can be used to create high resolution density output over Gpc/h box-sizes from low resolution simulations with negligible computational effort.}


\begin{document}
\maketitle
\flushbottom

\section{Introduction}

Cosmological galaxy surveys, both current and planned, span increasingly large volumes, eg. DESI \cite{desi}, EUCLID \cite{euclid}, DES \cite{des3} and LSST \cite{lsst}. In the previous decade, surveys such as the Baryon Oscillation Spectroscopic Survey \cite{boss} have used bright red galaxies as tracers of large scale structure. To make larger maps, the current and next generation of surveys rely on utilizing fainter galaxies. Spectroscopic surveys specifically are planning to observe galaxies targeted by the presence of star formation within them. Such galaxies, some of them Emission Line Galaxies (ELGs;  \cite{Raichoor_2017}), are the cosmological workhorse of the next decade. 
An important issue with creating bigger large-scale structure with smaller galaxies is being able to simulate such maps computationally. ELGs reside in halos at or below the mass scale of our own Milky Way \cite{Guo_2019, Gonzalez_Perez_2017}. Thus, simulations of such galaxy samples require not just large volume but also higher mass resolution than previously needed. A single simulation with equal volume of the DESI ELG sample that properly resolves all halos in which ELGs could form would require upwards of $9500^3$ particles, assuming a volume of 122 Gpc and a resolution of halos at around $1000$ particles for $10^{11} \ \textup{M}_{\Sun} $ halos. We often need a large suite of such simulations to explore various parameters or build up statistics. This will quickly become computationally expensive.

A promising solution to this problem is the use of deep learning networks. Various works in recent years applied convolutional neural networks to cosmological applications successfully, generally increasing small scale structure information (eg. galaxy distributions in \cite{Yip2019}, mapping dark matter to galaxies in \cite{kasmanoff2020dm2gal,Yip2019}, extracting cosmological information while marginalizing over baryonic effects in \cite{villaescusanavarro2020neural}, or for general simulation output \cite{He2019}, \cite{deoliveira2020fast} and \cite{Kodi_Ramanah_2020}). In the case of creating cosmological simulations spanning large volumes, a simple first approach is to use DM-only density fields.

In this paper, we propose a novel method to predict the $z=0$ density field of a high resolution simulation simply by giving it the $z=0$ density field of a low resolution simulation of the same volume, run using same cosmologies and initial conditions. High and low resolution here refers to the simulation particle's mass and the number of particles used inside the simulation volume. Once a network is trained successfully it takes little computational effort to produce new output from low resolution density fields. This would allow one to run a computationally less expensive, lower resolution simulation spanning Gpc/h in volume, and then using the resulting density field to predict the structure at higher resolution using the CNN. The halos and subhalos in this high-resolution output can be populated with galaxies in any number of ways, as required by the specific goal of the mock galaxy catalog (see, \cite{TinkerWechsler} for a review on the galaxy-halo connection).
This paper provides a first proof of work that uses density fields to super resolve dark matter halos.
Previous works either focus on other statistics such as void abundance \cite{Kodi_Ramanah_2020}, or use particle displacement fields that assumes lattice pre-initial conditions (pre-ICs, the Lagrangian particle positions) \cite{YinLi2021} and therefore cannot be applied to many of the state-of-the-art simulations that use glass pre-ICs.

In particular, our presented method here can be applied to dark matter simulations with glass initial conditions \cite{white1996} (here after: IC)  while previous super-resolution techniques such as \cite{YinLi2021_2} cannot. Many state of the art simulations use glass IC, such as Illustris project \cite{Vogelsberger2014b} while there is currently no satisfactory way to interpolate those glass IC particles onto a grid in order to compute the displacement field needed by previous super-resolution methods such as \cite{YinLi2021}. Any kind of interpolation technique will always create voxels that contain 0 particles, thus making displacement field calculation impossible for that pixel. 
Rather than wasting computational resources on new costly simulations for super-resolution tasks,
we develop the alternative Eulerian method to exploit the existing state-of-the-art cosmological simulations that started with glass IC.

\section{Method}

\subsection{Dataset}
\label{sec:dataset}

In this work, we used redshift=0 data from the Illustris-2-Dark (high-res) and Illustris-3-Dark (low-res) dark matter only simulations \cite{Vogelsberger2014b} for training and testing.
Illustris-2-Dark has a box size of $75$ Mpc/h, and contains $910^3$ ($m_{\text{dm}} = 6 \times 10^7 \ \textup{M}_{\Sun} $, high-res) and Illustris-3-Dark contains $455^3$ ($m_{\text{dm}} = 4.8 \times 10^8 \ \textup{M}_{\Sun}$, low-res) dark matter particles.
This results in both simulation boxes having the same mean mass density, meaning if more particles are simulated in the higher resolution simulation (Illustris-2-Dark in this case) they have to be lighter.
The cosmology for these simulations is consistent with WMAP-9 measurements \cite{Bennett_2013} and the glass tiled initial conditions are described in detail in \cite{Vogelsberger2014a}. Similar work done by \cite{YinLi2021} used displacement fields created from particle positions for training, and resolved only higher mass halos (lowest physical halo mass resolved at $z=0$ is $10^{11} \ \textup{M}_{\Sun}$,  containing 120 tracer particles). However, displacement fields are only naturally defined for lattice initial conditions, and are difficult to produce from glass tiled initial conditions, making this work ideal for simulations that do not have initial conditions that can be turned into lattice like configuration. 

Most state of the art hydro simulations (like Illustris) use glass ICs to improve statistics at high redshift snapshots, thus the underlying DM-only simulations are usually initialized using glass tiles, which are simply copied next to each other to fill the simulation volume. There is no simple way to interpolate those initial particles onto a grid in order to then use the displacement field super-resolution method pioneered in \cite{YinLi2021}. Any kind of interpolation technique will always have some residual pixels contain zero or 2 particles. Finding a general, automatic, way to fill those remaining pixels with exactly one particle in a coherent way is very impractical if not impossible, and has so far not been done. Additionally using displacement fields also fixes the number of pixels a given simulation has to be divided into. If a simulation contains for example $512^3$ particles, exactly one pixel per spatial coordinate and particle needs to be created for training and testing, complicating general applicability on any given simulation.

In our proposed method, we do however use grids to create density grids, but we do not need to create a displacement field (which requires having at least one particle per pixel in order to find the displacement field). 
Our deep learning method uses convolution layers, which are typically applied imaging data that has translational symmetry.
Convolutional neural networks (CNNs) generally work well with equidistantly gridded n-dimensional data. In our case, a fixed spatial resolution is perfect for a convolutional neural network. 
We prepare dark matter density maps, with a $2048^3$ grid, creating a number count density field where each voxel of side-length $\approx 36.6$ kpc/h contains information about the number of particles that lie inside that small region.
Each of the data sets is then subsequently divided into eight sub-cubes of equal size (37.5 Mpc/h on the side), from which six are used for training, one for validation, and the last one for testing. This results in the low resolution input (Illustris-3-Dark) and high resolution target (Illustris-2-Dark) simulations being divided into $32768$ 3-dimensional $64^3$ pixel cubes, of which an eighth is used for testing.

\subsection{Data pre-processing}
\label{sec:preprocessing}

Neural networks benefit from certain input and target data distributions more than others, in particular, they should not have a very long tail or an extremely large dynamic range. This makes the following pre-processing steps necessary for a successful and efficient training process.

\subsubsection{Mesh painting}

To achieve a data PDF shape which is continuous and roughly resembles a Gaussian distribution, the Illustris simulations' catalogs are painted onto a mesh using a TSC (Triangular Shaped Cloud) window kernel, since most voxels are empty, this helps to "smooth" out the integer number count values especially in low density voxels which importantly helps smoothing out the individual training cubes' PDF curves and removes large discontinuities.

\subsubsection{Normalization}

The ranges of feature distributions vary a lot among different training cubes and are also dependent on the training cube's position inside the simulation. To speed up and stabilize the training process, we normalize all training cubes as we, for example, do not want to learn to predict certain features better than others.
The following simple normalization $N$ is used:
\begin{equation}
N_{\text{log1p}}(x) = \text{log}\left(\epsilon + 1 + x\right),
\end{equation}
where the not strictly necessary factor $\epsilon$ is set to $\epsilon \equiv 10^{-7}$.

\subsubsection{Cube division}

The simulations painted on 3D meshes are too big to fit in a single GPU memory all at once. We then choose to use sub-volumes with $112^3$ voxels
as the input low-res data to the networks, and $64^3$ as the target high-res data.
These include periodically padded regions of width of 24 voxels,
and require no further ``paddings'' once inside the networks. 
By doing so we fully preserve the translational symmetry, which is crucial for generating outputs without artifacts near their boundaries.
Using the model described in \ref{sec:models and training}, this results in a generated output cube size of $64^3$ voxels.

\subsection{Models and training}
\label{sec:models and training}

At the $36.6$ kpc/h pixel scale, dark matter in the low and high resolution simulations are already spatially shifted from each other, due to the addition of the high-frequency modes in the high-resolution simulation.
The high density regions can shift by a few pixels.
This makes the simpler training approach of supervised learning (e.g.\ using a mean squared error loss function) impossible as the model does not have enough information 
to predict this shift and will simply blur the output.
Recent works (e.g.\ on displacement fields not density fields in \cite{YinLi2021}) showed that an unsupervised approach to deep learning is a possible pathway forward given the limitation described here.

First proposed by Ian Goodfellow in 2014 \cite{Goodfellow2014} Generative Adversarial Networks (GANs) were introduced as a new
class of generative models.
GANs are able to produce output from input sampled from a random noise distribution and generally do not rely on information from the input data to make predictions.
GANs consist of two different networks that are trained in parallel to each other: the generator G and the discriminator D.
For a given set of real training samples, the generator learns to produce fake samples that look similar to the real ones, while the discriminator tries to learn to differentiate between real and fake samples. Formally (as in \cite{Goodfellow2014}): we can define a mapping to data space $G(z, \theta_g)$ where $z$ is taken from a prior, and usually random, distribution $p_z(z)$, whereas $\theta_g$ are G's learnable parameters. Similarly, we define a mapping $D(x, \theta_d)$ that outputs a single scalar representing the probability that $x$ was taken from the data distribution ($p_{\text{data}}$) and not from $p_g$ (the "fake" distribution produced by G). Training both D and G means maximizing the probability of D correctly identifying fake and real data, while simultaneously training G to minimize $\log \left(1 - D(G(z) \right) $. This means that D and G play a minmax game with value function:
\begin{equation}
\begin{aligned}
\min_G \max_D V(G,D) = \ &\mathbb{E}_{x \sim p_{\text{data}}(x)} \left[ \log D(x) \right] + \\
 &\mathbb{E}_{z \sim p_z(z)} \left[ \log (1 - D(G(z))) \right],
\end{aligned}
\end{equation}
where $\mathbb{E}$ denotes the expectation value, which in practice is replaced by the
average over a batch with stochastic gradient descent.
As both D and G improve in their training, the desired end result would be that G produces fake samples that D is no longer to differentiate, thus having a 0.5 probability to guess correctly.
This state is also called Nash-equilibrium \cite{Nash1950}. Of course, the training of actual models will stop before this happens, but if the training was stable, G might be trained well enough to produce valuable results before Nash-equilibrium is reached.

Nevertheless, in this work G needs to generate output not from a randomly sampled input distribution, but from the low resolution training cubes and from white noise fields in the noise layers. In order to help the discriminator differentiate between real and fake data, the low-res input is concatenated to the generated output and  high-res target before passing the data through D. The training process thus is no longer purely unsupervised, and our method is a conditional GAN, an adapted version of the standard approach first introduced by \cite{mirza2014}, which utilized a different architecture.\\

This results in the following loss functions for the discriminator and generator in this work:

\begin{align}
\begin{aligned}
\mathcal{L}_D = \ &-\mathbb{E}_{\hat{x},x} \left[ \log D(\hat{x},x) \right] \\
&- \mathbb{E}_{\hat{x}, z} \left[ \log (1 - D(\hat{x},G(\hat{x}, z))) \right] \\
&- \gamma \ \mathbb{E}_{\hat{x},x} [  \left\Vert \nabla D(\hat{x},x) \right\Vert^2 ]
\label{eqn:d_loss}
\end{aligned}
\end{align}
\begin{align}
\begin{aligned}
\mathcal{L}_G = \ &- \mathbb{E}_{\hat{x}, z} \left[ \log (D(\hat{x},G(\hat{x},z))) \right],
\end{aligned}
\end{align}\\
where $x$ is sampled from the high-res distribution $p_{\text{high-res}}$, $\hat{x}$ is sampled from the low-res distribution $p_{\text{low-res}}$ and $z$ is sampled from a random white noise distribution $p_z$. The third term in equation \ref{eqn:d_loss} is the $R_1$ regularization (introduced in \cite{Mescheder2018}), which is applied for real data only. The chosen penalty weight $\gamma=5$ is constant.
\subsubsection{Models}

\begin{figure}
\centering
	\includegraphics[width=0.8\columnwidth]{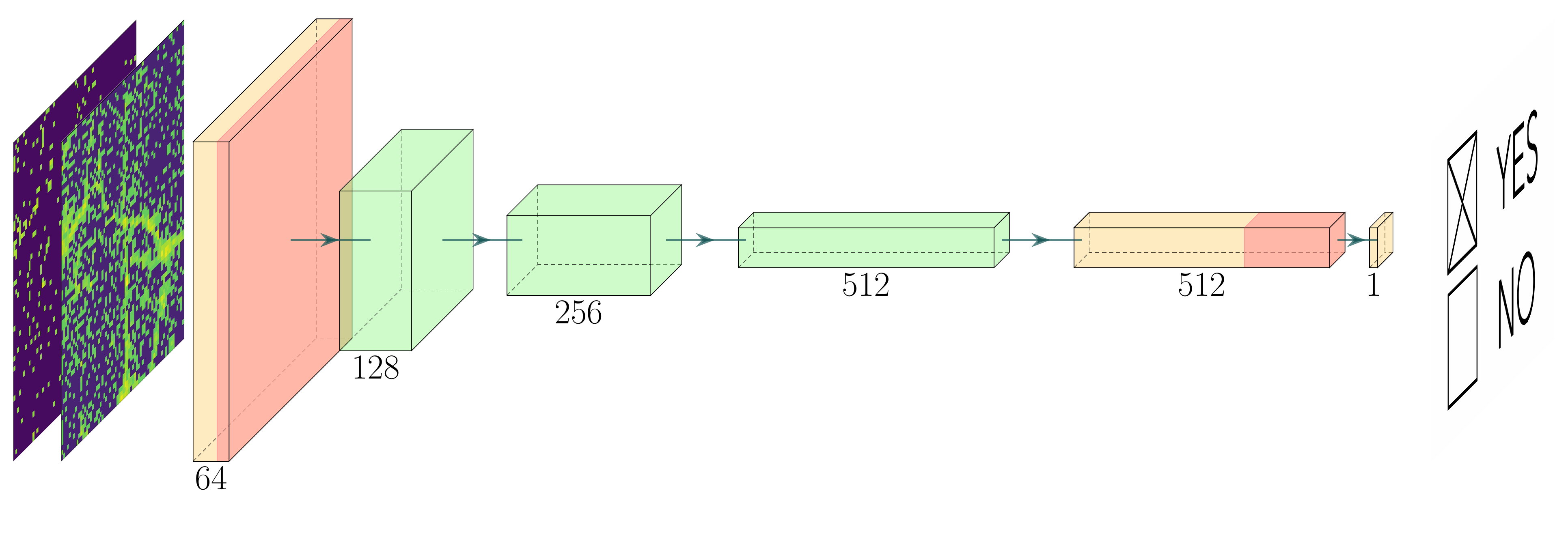}
    \caption{A plot showing the rather simple architecture of the discriminator used for GAN training. Of course, every stage applies 3D operations on a 3D input, but the visualization works better in 2D. The net's stages are indicated as: \textit{light orange}: a 3D convolution; \textit{red}: LeakyReLU activation; \textit{green}: Residual Block: consisting of two 3D convolutions each followed by a LeakyReLU activation, another 3D convolution and a linear down-sampling layer. At the end D outputs a $4^3$ voxel cube which is averaged over before being passed to a sigmoid inside the BCELoss function. For further details please refer to the data availability section.}
    \label{fig:discrim}
\end{figure}

\paragraph*{Discriminator:}
Training is not too sensitive on the exact discriminator architecture, as long as the main idea of progressively extracting higher-level features by down-sampling convolutions is followed. Figure~\ref{fig:discrim} depicts a qualitative approximation of the architecture as not all shapes and sizes are completely accurate. It is important to understand that each convolution changes the shape of the output feature map slightly, but since this would barely be visible in the graph it is neglected here. Additionally, in this case the boxes indicate the operator acting on the feature map of each stage. Meaning that the actual data being passed through the network follows the path indicated by the arrows and is reshaped by the operators along the way. For the exact model architecture, please refer to the data availability section.

\begin{figure}
\centering
	\includegraphics[width=0.8\columnwidth]{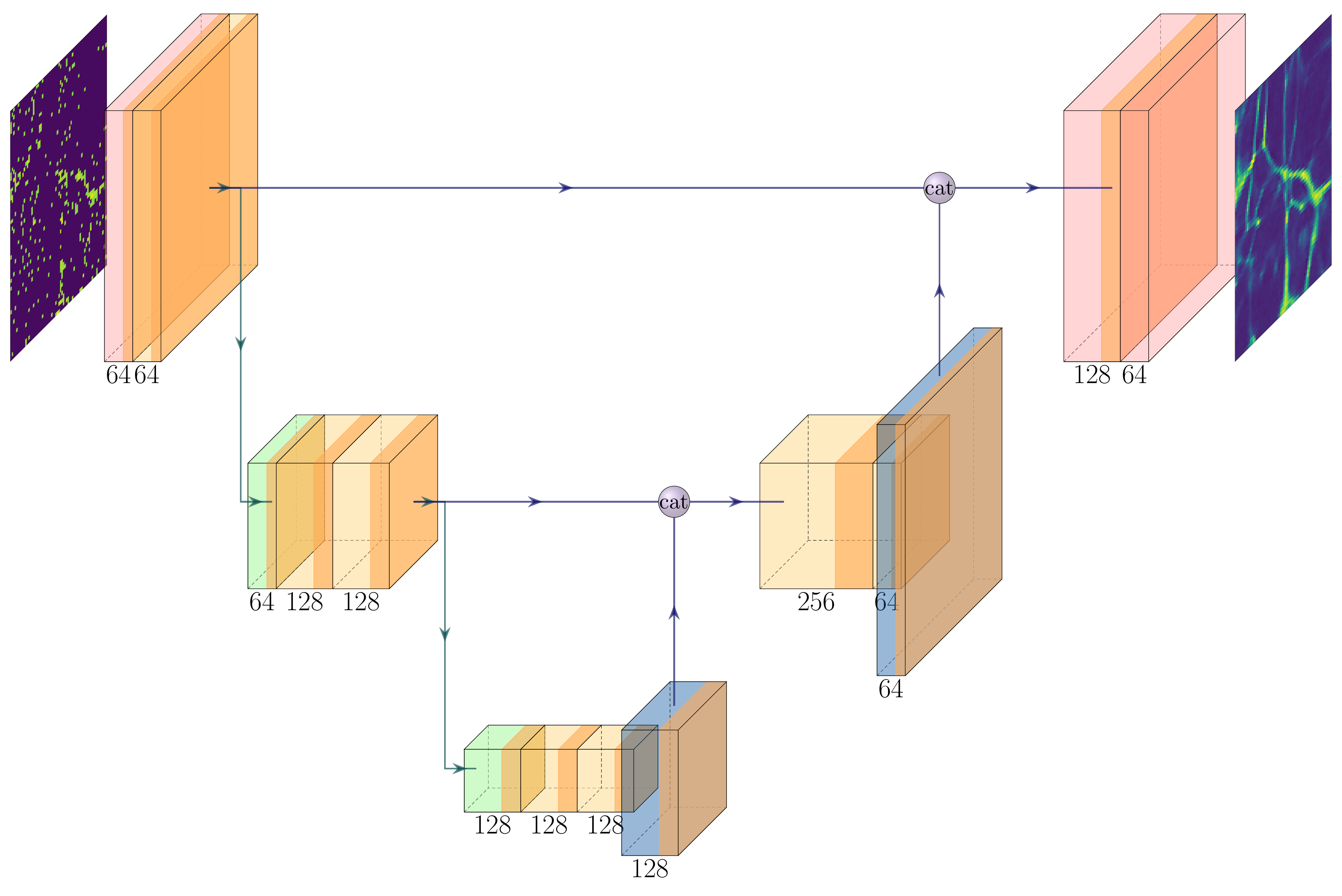}
    \caption{The U-Net used as a generator in this work. Each stage's output's number of channels is depicted by the small number below. The spatial dimensions of the feature maps at each stage are approximately depicted by the width and height of each stage. The net's stages are indicated as (different colors as for D in figure \ref{fig:discrim}):  \textit{light red}: a 3D convolution; \textit{light orange}: a noise layer followed by another 3D convolution; \textit{dark orange / brown}: LeakyReLU activation; \textit{green}: noise layer followed by a 3D (down-)convolution; \textit{blue}: noise layer followed by a linear up-sampling layer, a 3D convolution, a LeakyReLU activation, another noise layer, another 3D convolution and lastly another activation; \textit{cat}: concatenation of two outputs from two layers along the channel dimension. For more exact details please refer to the data availability section.}
    \label{fig:gen}
\end{figure}

\paragraph*{Generator:}
The model's exact architecture is extremely important to a successful training process. The model, itself not too complicated or large, is a shallower adaptation of the often used U-Net (introduced in \cite{Ronneberger2015}, adapted in eg. \cite{Giusarma2019,He2019} and \cite{chen2020learning}). As explained in \cite{Shelhamer2017}, convolutional neural nets are built on translational equivariance since their component stages only depend on relative spatial positions as they operate on local input regions. Thus at all blocks each voxel's receptive field has the same size, independent on how large or small the initial input is. This translational equivariance is very important as it ensures that the net will produce good output on a variety of different inputs. The most important change from a conventional U-Net is that we replace the transposed upsampling convolution by a tri-linear interpolation followed by a usual convolution (detailed implementation in figure \ref{fig:gen} and the data availability section, explained further in \cite{odena2016}). This greatly reduced the checkerboard effect produced by the transposed convolutions, thus helping the training process immensely. \\
In order to help the generator predict the highly non-linear structure of the high-res target, uniformly sampled random noise is added before each convolution. This provides the basis upon which the generator can build its small scale prediction on. Similar to the hierarchical structure formation of dark matter, the generator's U-Net structure causes the layers producing small scale structure to be conditioned on the large scale features that are passed through G. This inherent similarity makes U-Nets an obvious choice for working with cosmological simulation data.

\subsubsection{Training}
\label{sec:training}

Each epoch of the training process was divided into training with the training data set, followed by validation on the entire validation data set. To increase the training set's distribution variability, we augmented the dataset by flipping and permuting axes of the training cubes. This helps the network learn discrete rotations and parity symmetries which are inherent in cosmological simulations. Such augmentation was not done on the validation set for deterministic results at every epoch.
In the first epoch of training, only the generator was trained using the standard supervised approach using MSE Loss. This helped the generator learn large scale shapes of the low-res input and high-res target which was beneficial to the later epochs of GAN training.
With the AdamW optimizer \cite{adamw2019} and small weight-decay ($\lambda = 10^{-4}$), we adopt different learning rates for the two networks: $1 \times 10^{-5}$ for the discriminator and $5 \times 10^{-5}$ for the generator, with a ratio similar to the one suggested in \cite{ttur2017}.
It is important to notice that these learning rates are not universally applicable and will need to be adjusted for different training applications to reach an equilibrium training state. Furthermore a $R_1$ regularization gradient penalty scheme \cite{Mescheder2018} was used every 16 batches to penalize the discriminator upon deviating from equilibrium on real data alone, thus increasing training stability.

\section{Results}

The sections below will summarize the quality of the network's output after training it for 600 epochs using the method described in section \ref{sec:training}. Not every epoch produces equally good results, although at later stages of training the output quality is stable. This fluctuation is due to both networks (D and G) hovering around a local loss minimum, only incrementally improving. Most statistics are more or less unaffected by this, but for example the halo two-point function for small mass halos is very sensitive to the exact output. The best performing epoch of the last 10 is selected, in this case: epoch 599. Best performing in this case refers to good accordance with all target statistics. Epoch 599 was not the only one producing good results across all statistics, but was subjectively selected for providing the best results. It should also be mentioned that extending that run further until epoch 1200 did not increase the quality of the output as the training seemed to hover around a local minimum.

The testing dataset consists, as described in section \ref{sec:dataset}, of one eighth of the entire simulation region. Thus all testing cubes combine, stitched back together, a volume of $37.5$ Mpc/h. All plots will show comparisons between the Illustris-3 low-res input, the generator's output and the high-res Illustris-2 target, labeled as \textit{low-res}, \textit{generated} and \textit{high-res}.

Some statistics (eg. two-point function, see below) require the data to be stored as particle catalogs and not as number-count-field pixel values. In order to go back to catalog format in the most simple and physical way, two steps are taken:

First, all voxels are Poisson sampled as follows:
\begin{equation}
x_\text{poisson} = \begin{cases} x &\text{if} \ r > x -  \floor{x}\\
x+1 & \text{else},
\end{cases}
\end{equation}
where $r$ is uniformly drawn between $0$ and $1$. This results in the net output's voxels being converted to integers only. Secondly, if a voxel's Poisson sampled value is eg. 4, four particles are randomly placed inside the voxel's volume and depending on that voxel's relative position inside the simulation, a position-catalog entry is made for each particle.

This method obviously completely erases any information about particle positions below the pixel-size scale, but this information is already lost upon creation of the training data. Thus this kind of pixelization poses fundamental limits and challenges to using number count or density fields for this training method, as compared to e.g.\ displacement field training (which however does not apply to simulations with glass initial conditions such as Illustris here).
This effect will also become more apparent in section \ref{sec:stats}, while we look at statistical properties of the output.

\subsection{Visual Comparison}

\begin{figure*}
	\includegraphics[width=\textwidth]{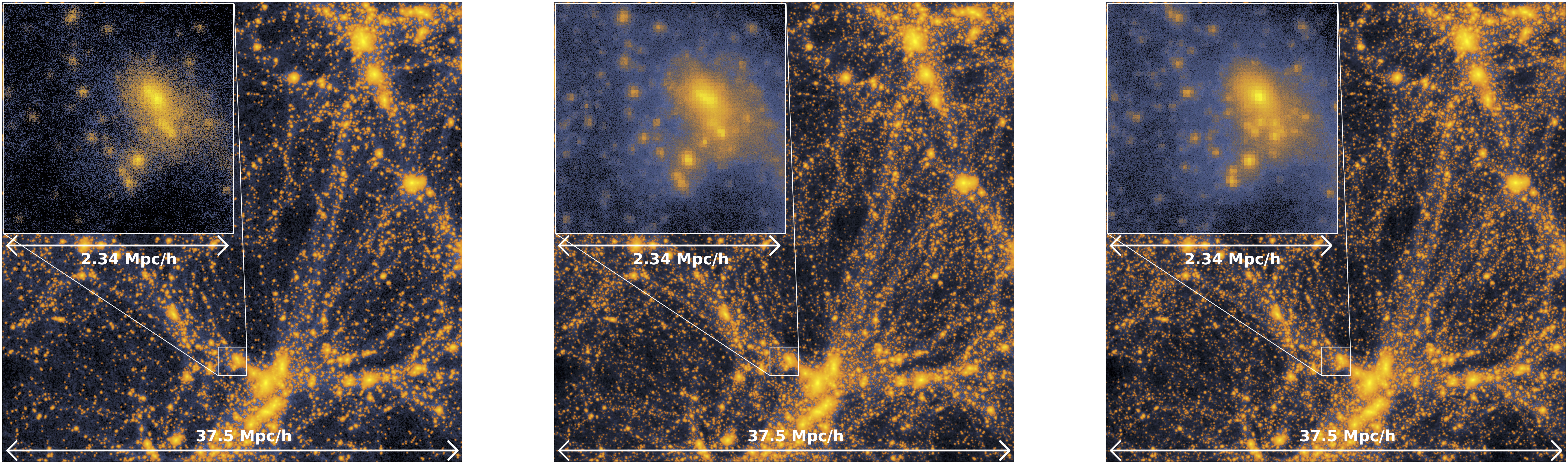}
    \caption{\textit{left:} low-res catalog, \textit{middle:} generated output catalog, \textit{right:} high-res catalog. A projection on the z-axis of each catalog's particles, where particles which are part of FoF halos, are highlighted in orange. The generated output plot is indistinguishable from the high-res plot by eye on large scales. At smaller scales, the shapes and positions of small halos look statistically consistent but vary a bit between the two as expected, whereas they are completely missing in the low-res plot. The depicted testing-region box is not periodic as it only spans one eighth in volume of the entire corresponding Illustris simulation, which spans $(75 \, \textrm{Mpc}/h)^3$.
    This figure was created with the help of Yueying Ni.}
    \label{fig:projection}
\end{figure*}

The visual differences between the low-res input and generated output / high-res target simulation become clear immediately in figure \ref{fig:projection}, especially when looking at the smaller mass halos that are missing in the low-res simulation data.
As expected, the generated output match the high-res simulation on large scales thanks to the conditional GAN training, while on small scales the generated fine structures are different from that in the simulated ones but look statistically consistent.
For quantitative comparison, below we look at various matter and halo statistics which provide a more solid and consistent way to quantify model performances.

\subsection{Statistics}
\label{sec:stats}
There are numerous different statistical properties that can be looked at, but most of them, eg. the power spectrum, don't provide any new information, as both the low-res input (Illustris-3) and high-res target (Illustris-2) simulation are already at a quite high resolution and thus are non-differentiable at this pixel-size scale. But, crucially, some statistics do show clear differences, proving the performance of the trained network.

\subsubsection{Power spectrum and 2PCF}
\begin{figure*}
\centering
	\includegraphics[width=0.8\textwidth]{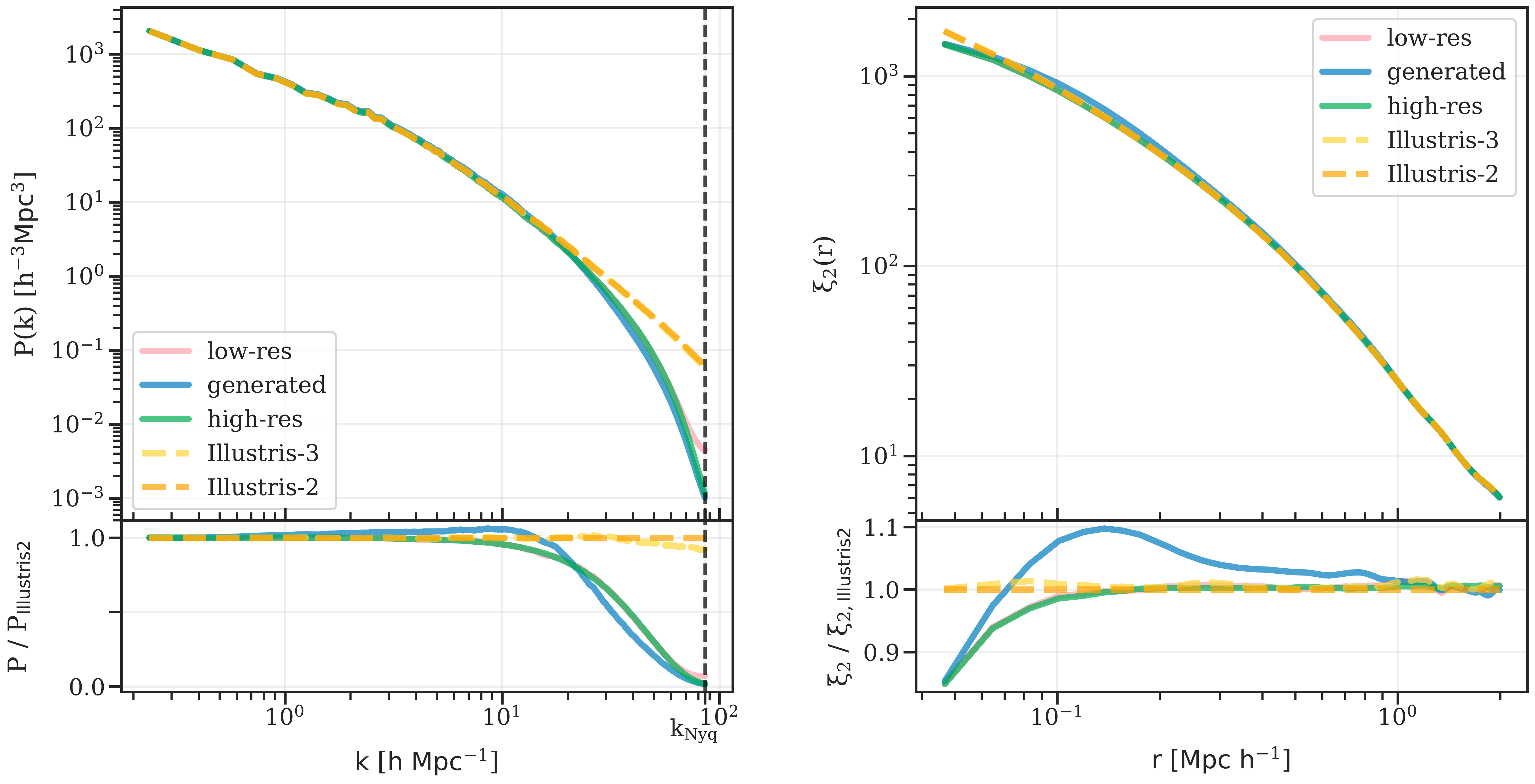}
    \caption{\textit{left}: The particle power spectrum comparison between the catalogs obtained from Poisson sampling the training data and the "true" (unpixelized) Illustris simulation catalogs. It is important to also compare to the "true" Illustris statistics, as this will give an idea of the effects the pixelization has on the low and high-res testing data. The divergence at small scales due this pixelation becomes clearly visible. The low-res and and high-res simulation's power agree extremely well with each other up to the Nyquist frequency, thus making the comparison futile as a way to prove the neural net's performance. \textit{right}: The particle two-point correlation function up to bin sizes of 2 Mpc/h, calculated using the Landy-Szalay estimator. As with the power spectrum, the low and high resolution simulations coincide at scales above the pixelsize of ${\sim}36.6$ kpc/h, only proving that the network can at least mirror the input roughly in that regard. Small differences between low-res input and generated output originate from the fact that the net does not simply learn to copy the low-res input. The "true" Illustris two-point correlations differ slightly from the testing data as an effect of the pixelation.
    }
    \label{fig:power_2pcf}
\end{figure*}

Figure \ref{fig:power_2pcf} depicts the particle 3D power spectrum and particle auto-correlation function of the entire testing box, defined as in for example \cite{movandenboschwhite2010}. They are compared to the power spectrum and two-point function calculated from the "true" Illustris simulation particle catalogs over the same region. The power spectrum $P(k)$ is calculated using a FFT over a cic-painted meshgrid whereas the two-point function $\xi_2(x)$ is calculated by counting particle pairs in binned distance-regions.
\begin{align}
	\xi_2(x) &\equiv \left\langle \delta(\bm{x_1})\delta(\bm{x_2}) \right\rangle, \quad x = |\bm{x_1} - \bm{x_2}|\\
	\xi_2(x) &= \frac{1}{(2\pi)^3 } \int d^3k \ P(k) e^{ikx},
\end{align}
where the power $P(k)$ is simply defined as the Fourier transform of the two-point correlation function and $\delta(\bm{x_i}) = \frac{\rho(\bm{x_i})}{\bar{\rho}} - 1$ is the matter overdensity. The effect of cic-compensation (as described in eq 18 of \cite{Jing_2005}), which aims to reduce the statistical effect that painting particles onto a mesh has on the power, is negligible and does not reduce the difference between the pixelated and true power spectra.

For both, power and two-point function, the output agrees very well with the low-res input / high-res target. The power differences at small physical scales between testing data / generated output and the true simulation catalogs, are caused by the pixelation and the ensuing loss of information below pixel-size scales. The autocorrelation plot only covers a smaller range of distances, thus doesn't show this diversion. Since both the power and two-point function of input and target are equal at these scales, it comes as no surprise that the output will match these statistics reasonably well. It should also be stated that even though the low-res input distribution matches the high-res target's in these two statistics, the generator does not produce them exactly, as it does not simply learn to produce a one to one mapping between input and target.

\subsubsection{Halo mass function}
\begin{figure}
\centering
	\includegraphics[width=0.5\textwidth]{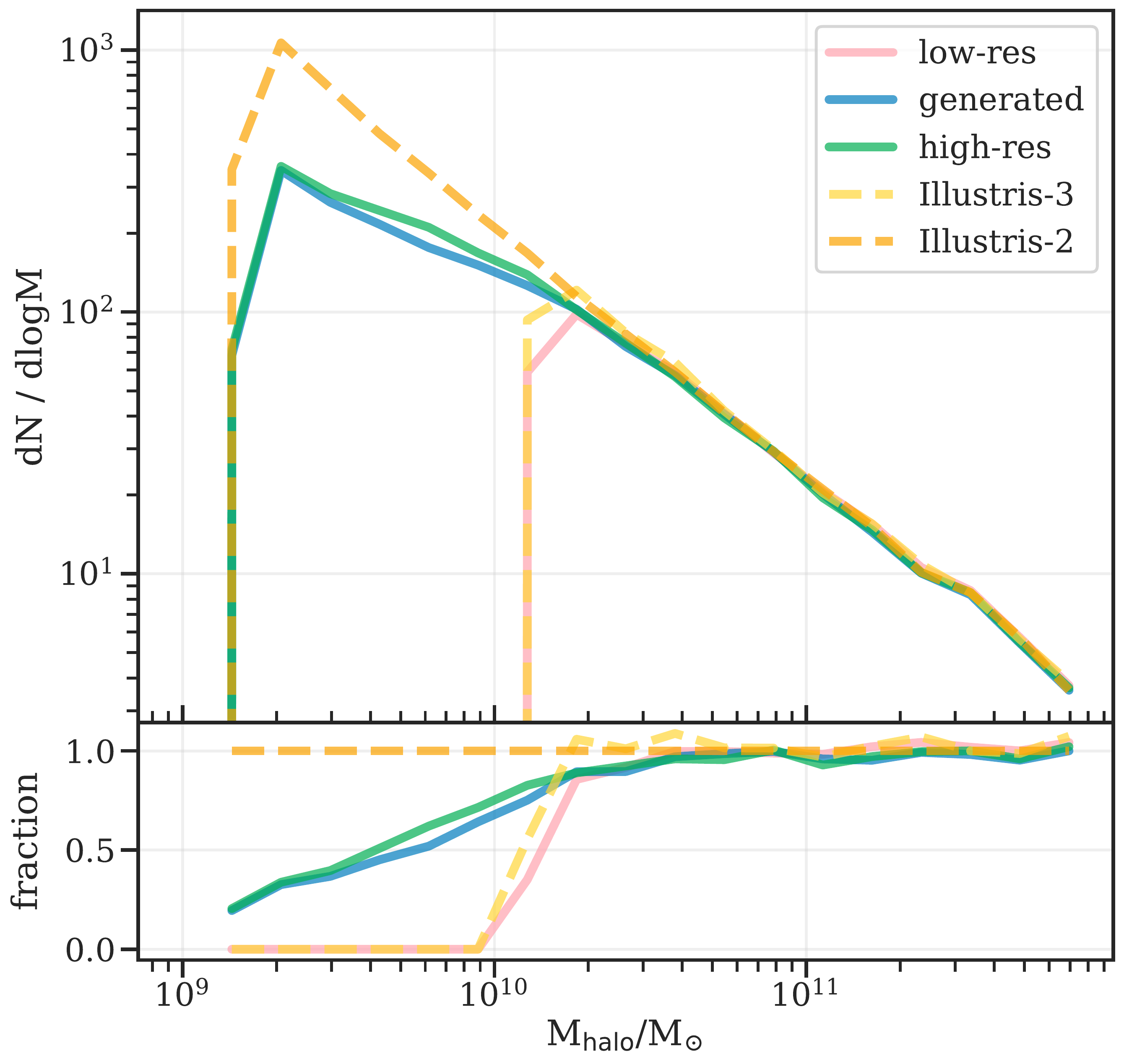}
    \caption{The halo mass function obtained from FoF halos (linking length $0.2$) inside the testing region. Illustris halos are taken from the "true" high- and low-res $z=0$ simulation snapshots, which importantly show the effect the randomization of particle positions inside each voxel has on FoF halo finding. This effect becomes clearly visible for halos below $10^{10}$ M$_{\Sun}$ as the output and target halo mass functions start to diverge from the "true" Illustris-2 halos, whereas the "true" Illustris-3 halos still coincide with the halos found from the low-res testing data as they are more massive and thus unaffected by the pixelation.}
   \label{fig:hmf}
\end{figure}

All halos in figure \ref{fig:hmf} where calculated using nbodykit's \cite{nbodykit} FoF (Friends of Friends) halo finder using a relative linking length of $0.2$. Again, the pixelation effect becomes visible at smaller mass halos below and around $10^{10}$ M$_{\Sun}$, as the HMF found from pixelized data starts to deviate from the halos found using the true simulation catalog. At smaller/larger pixel sizes this effect will shift to smaller/larger halo masses.

\subsubsection{Halo two-point function}
\begin{figure*}
\centering
	\includegraphics[width=0.8\textwidth]{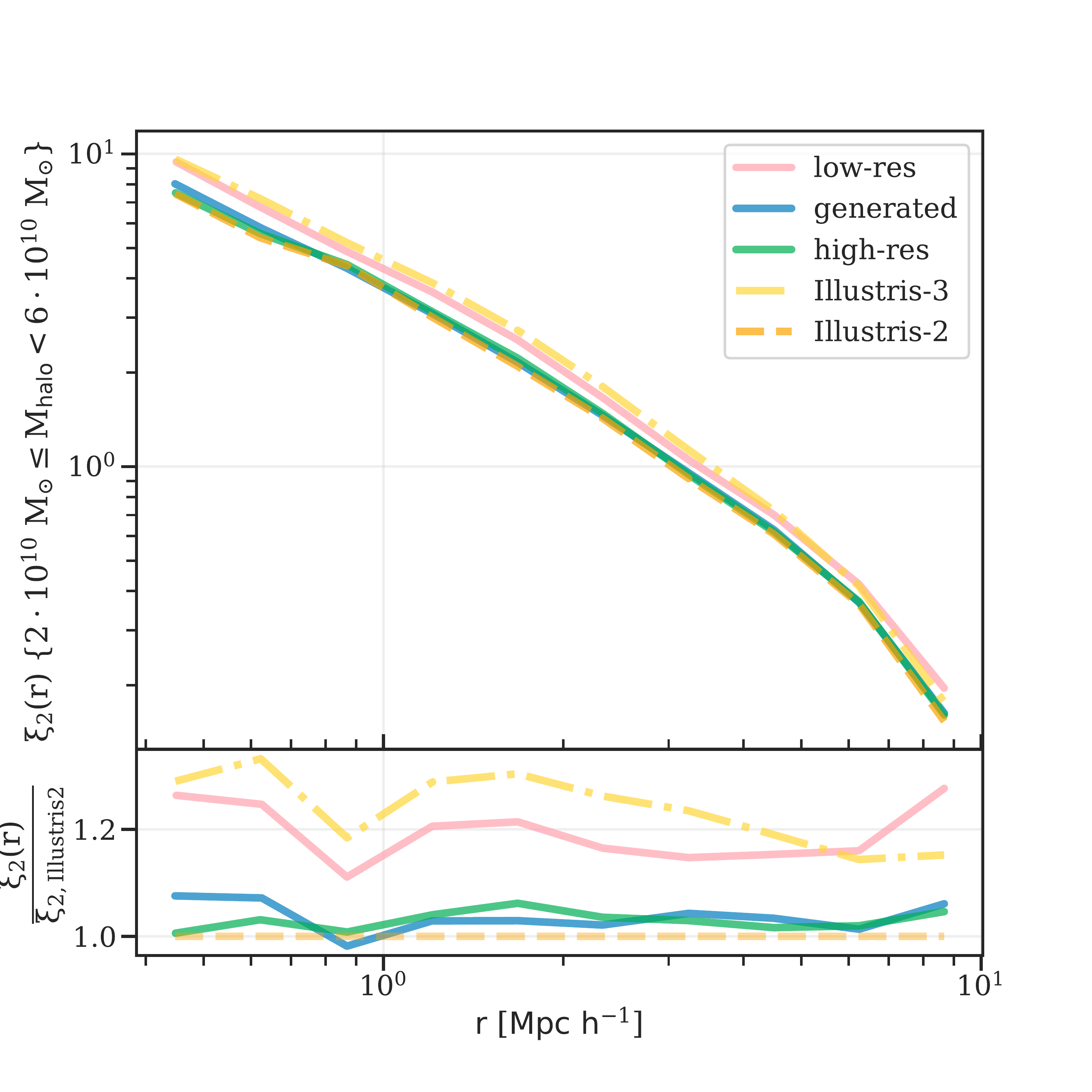}
    \caption{The halo two-point function comparison between the testing data and the "true" un-pixelated Illustris simulation FoF halos of the testing region, calculated using the Landy-Szalay estimator. Only halos between $1 \times 10^{10}$ M$_{\Sun}$ and $6 \times 10^{10}$ M$_{\Sun}$ are plotted. The network correctly predicts the high-res target's (and also "true" Illustris-2) correlation of halo positions, which importantly differs from the low-res input's. This plot is the most critical one, as it indicates that the method learns to predict the high-resolution simulation and not simply up-scales the low-resolution density fields. It also becomes clear from the comparison to the "true" Illustris halo two-point correlation functions, that pixelation only has a minor effect on the testing data in this regard.}
    \label{fig:halo_2pcf_1}
\end{figure*}

\begin{figure*}
\centering
	\includegraphics[width=0.8\textwidth]{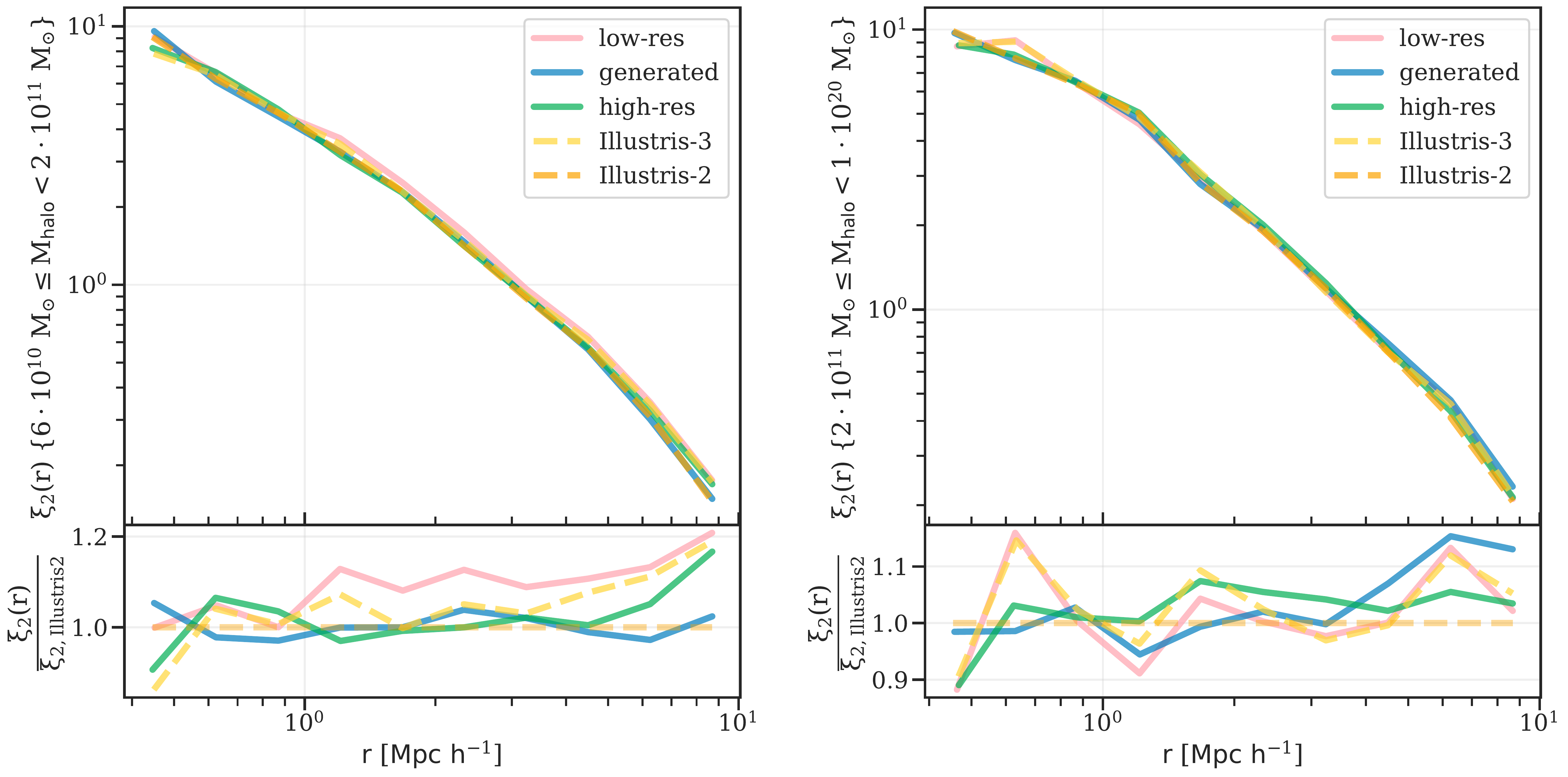}
    \caption{Halo two-point functions of higher mass bins. Here low-res input and high-res target start to coincide again so the net's performance becomes less meaningful at these mass scales.}
    \label{fig:halo_2pcf_2}
\end{figure*}

The halo auto-correlation function is calculated by using the halo's central positions. Figure \ref{fig:halo_2pcf_1} and \ref{fig:halo_2pcf_2} shows the two point function of different halo mass bins. Importantly for halos between $1 \times 10^{10}$ M$_{\Sun}$ and $4 \times 10^{10}$ M$_{\Sun}$, a clear difference between the low resolution and high resolution simulation can be seen. The network manages to match the high resolution simulation perfectly, proving it's performance. It also matches the two-point functions for higher mass halos, which is to be expected as those halos are larger in size and easier to predict. The two-point function for halo masses below $10^{10}$ M$_{\Sun}$ becomes less meaningful, as the FoF halos found there will be affected due to the pixelation. Nevertheless the two point function of the (pixelized) generated output and high-res target still coincide with the two-point function calculated from FoF halos found by using true Illustris-2 simulation particles.

\subsection{Generalizing to another simulation}
\begin{figure*}
\centering
	\includegraphics[width=0.8\textwidth]{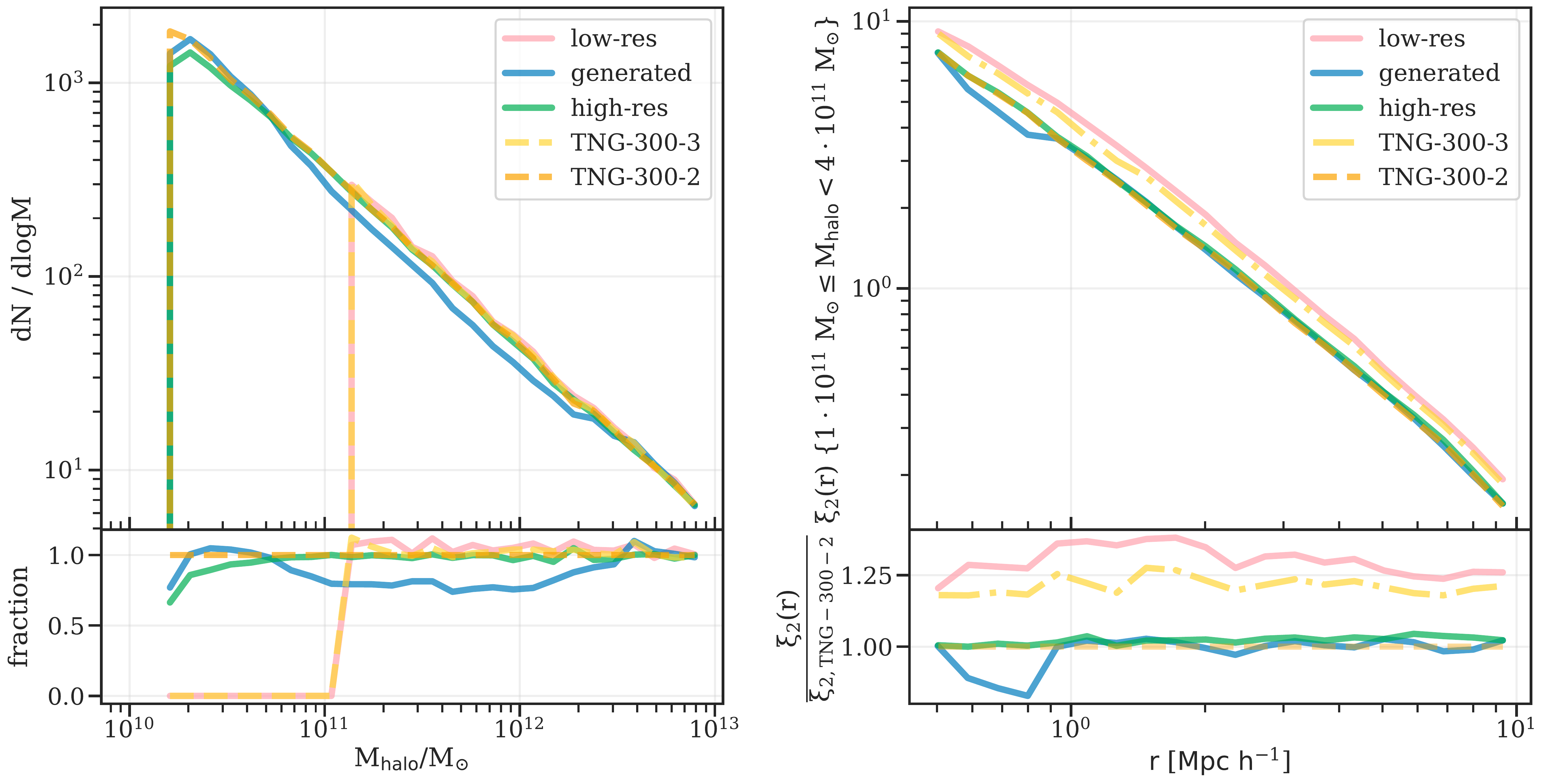}
    \caption{\textit{left}: The halo mass function of FoF halos of the TNG testing region, similar to figure \ref{fig:hmf}. Clearly the net performs weaker than on the Illustris data, but still manages to predict new lower mass halos, compared to the low-res input (TNG-300-3) simulation. Due to higher particle masses in the simulations, only higher mass FoF halos, compared to the Illustris halo mass function, are found. This also leads to the pixelation effect in figure \ref{fig:hmf} not yet to take effect, as it only affects FoF halos below $10^{10}$ M$_{\Sun}$ at this pixelsize. 
    \textit{right}: Halo two-point functions of halos between $1 \times 10^{11}$ M$_{\Sun}$ and $4 \times 10^{11}$ M$_{\Sun}$. Here low-res input and high-res target differ similarly to the Illustris simulations, but at higher halo masses, since the simulation's particle masses are different. The network makes the correct prediction without ever training on the TNG data set, indicating that it is applicable to a wide range of simulations. Again it becomes clear from the comparison to the "true" TNG-300 halo two-point correlation functions, that pixelation only has a minor effect on the testing data in this regard.}
    \label{fig:tng_halo2pcf}
\end{figure*}

Ideally, the above model would be perfectly applicable to any low resolution DM-only simulation's density field of similar pixel size and find new halos. The extend of this performance is tested in this section by using a set of the Illustris-TNG DM-only simulations \cite{Springel_2017}, which are the latest improvement on the previous Illustris simulations. They use almost the same cosmology, but of course are run using different initial conditions. By dividing both the TNG-300-3-Dark (low-res) and TNG-300-2-Dark (high-res) simulations into $5632^3$ density voxels, a similar pixelsize as before of ${\sim}36.6$ kpc/h is fixed for the data. For memory efficiency, only an eighth of the entire simulation volume of $205^3$ Mpc/h is used for testing again. Remarkably, even though the particle masses are different compared to the Illustris simulations (high-res TNG-300-2: $m_{\text{dm}} = 5.6 \times 10^8 \ \textup{M}_{\Sun} $, low-res TNG-300-3: $m_{\text{dm}} = 4.5 \times 10^9 \ \textup{M}_{\Sun} $), the network still manages to predict new halos successfully which can again be seen in both the halo mass function and halo two-point function in figure \ref{fig:tng_halo2pcf}. This is an important indication that convolutional neural networks are, once trained successfully, even applicable outside their initial training and testing region. Of course once the pixel size differs too much from the initial pixel size the network was trained on, deteriorating results are to be expected.

\section{Conclusions}

We apply GAN techniques to improve cosmological simulations to successfully generate high resolution outputs from low resolution counterparts, mimicking authentic high-res simulation results in all evaluated statistics (crucially in figure \ref{fig:halo_2pcf_1}).

This work pushes the resolution limit (from for example \cite{YinLi2021_2} and \cite{doogesh19}), while also being more generally applicable to general DM-only simulations than displacement field training solutions, as it only relies on knowing the particle positions at snapshot $z=0$. It provides an important first step to building a full suite going from low-res dark matter simulations, to ultimately producing ELG mock catalogs over Gpc in scale, by predicting new halos at mass ranges relevant to ELGs.

In the case of simulations where initial conditions are set using a glass tile configuration (like the Illustris simulations) it becomes almost impossible to gain information about the dark matter displacement field, as the initial particle positions cannot easily be placed on a grid together with their IDs.
This work allows to generate high-res density fields by only investing the CPU time it takes to run the low-res simulation instead of actually running the entire high-res simulation, as testing only took around 26 CPU hours for this work.
It is also naturally scaleable to much larger testing regions by simply providing a larger low-res input.

A natural downside to using density fields for training is mainly the fact that all information below pixel-size scales is lost naturally due to pixelization. This poses natural limits to the lowest mass of physical halos to be found when training on a given pixel-size. Nonetheless this work proves that the method was successful down to scales where the low and high resolution statistics start to deviate, and suggests that the same will also hold up for training on smaller pixelsizes. This will also lead to a better resolution of halos and subsequently also subhalos (identified using density information), which will lead to the initial goal of creating large scale high resolution ELG mocks from a low resolution simulation with almost no additional computational effort.

\section{Outlook}

Future improvements of this method might make use of a smaller pixel-size, eg. ${\sim}18.3$ kpc/h, to improve smaller mass halo predictions and enable subhalo predictions. In principle not much except for the learning rates will have to be adjusted, although the amount of training data will increase significantly, making the entire training process longer and more difficult to tune.
Additionally it will also be beneficial to train the network on multiple redshifts, to make it more generally applicable to DM-only simulation output.
Recently similar work using displacement fields and super resolution methods \cite{YinLi2021, YinLi2021_2} showed that training for higher redshift data is an easier task.
This also makes intuitively sense, as $z=0$ simulations went through the maximum amount of non-linear structure development, making it harder to learn the mapping accurately.

Another useful improvement to the training would be to add velocity fields to the input, containing information about the particles' bulk velocities inside each pixel-volume, helping to infer halo field phase-space distributions. The networks architecture supports training on multiple fields natively, so this change could be easily executed.

It will also be interesting to provide the net with completely different simulations, with different cosmological parameters and at similar mass resolutions and pixel-scales during training, to gain a more universally applicable network.

As we ultimately want to produce data comparable to observations, one will also have to infer redshift-space distributions from our real-space data, which will be subject of future work.

With the end goal in mind of creating a network that does not have to be retrained and can easily enhance low resolution simulation density output, this work provides an important foundation on the way to easily accessible large scale ELG mock catalogs.

\section*{Acknowledgements}

We thank Yueying Ni for sharing her plotting scripts.
This work was supported in part through the NYU IT High Performance Computing resources, services, and staff expertise. We would also like to thank AMD for the donation of critical hardware and support resources from its HPC Fund, that also made this work possible. The Flatiron Institute is supported by the Simons Foundation. JLT acknowledges support of NSF grant number 2009291.


\section*{Data Availability}

All code and models used and created in this work are publicly available on GitHub under the following url: \url{https://github.com/dschaurecker/dl_halo}. Please refer to the ReadMe file in the \texttt{dl\_halo} repository for a very in-depth guide through the code and the process of training and testing.
The training code builds upon the \texttt{map2map} code repository, available on GitHub as well (\url{https://github.com/eelregit/map2map}).
It allows for general training of $n$ arbitrary input fields to $n$ arbitrary output fields using custom models, normalizations and loss functions.
Furthermore some utilities from \texttt{nbodykit} an "Open-source, Massively Parallel Toolkit for Large-scale Structure" \cite{nbodykit} were used in pre-processing and statistical evaluation of this work.

\bibliography{paper} 

\providecommand{\href}[2]{#2}\begingroup\raggedright\begin{thebibliography}{10}

\bibitem{desi}
{DESI Collaboration}, A.~Aghamousa, J.~Aguilar, S.~Ahlen, S.~Alam, L.E.~Allen
  et~al., \emph{The desi experiment part i: Science,targeting, and survey
  design},  2016.

\bibitem{euclid}
R.~Laureijs, J.~Amiaux, S.~Arduini, J.L.~Auguères, J.~Brinchmann, R.~Cole
  et~al., \emph{Euclid definition study report},  2011.

\bibitem{des3}
{DES Collaboration}, T.M.C.~Abbott, M.~Aguena, A.~Alarcon, S.~Allam, O.~Alves
  et~al., \emph{Dark energy survey year 3 results: Cosmological constraints
  from galaxy clustering and weak lensing},  2021.

\bibitem{lsst}
Z.~Ivezic, S.M.~Kahn, J.A.~Tyson, B.~Abel, E.~Acosta, R.~Allsman et~al.,
  \emph{Lsst: From science drivers to reference design and anticipated data
  products}, {\emph{The Astrophysical Journal} {\bfseries 873} (2019) 111}.

\bibitem{boss}
K.S.~Dawson, D.J.~Schlegel, C.P.~Ahn, S.F.~Anderson, Ã.~Aubourg, S.~Bailey
  et~al., \emph{The baryon oscillation spectroscopic survey of sdss-iii},
  \href{https://doi.org/10.1088/0004-6256/145/1/10}{\emph{The Astronomical
  Journal} {\bfseries 145} (2012) 10}.

\bibitem{Raichoor_2017}
A.~Raichoor, J.~Comparat, T.~Delubac, J.-P.~Kneib, C.~Yèche, K.S.~Dawson
  et~al., \emph{The sdss-iv extended baryon oscillation spectroscopic survey:
  final emission line galaxy target selection},
  \href{https://doi.org/10.1093/mnras/stx1790}{\emph{Monthly Notices of the
  Royal Astronomical Society} {\bfseries 471} (2017) 3955–3973}.

\bibitem{Guo_2019}
H.~Guo, X.~Yang, A.~Raichoor, Z.~Zheng, J.~Comparat, V.~Gonzalez-Perez et~al.,
  \emph{Evolution of star-forming galaxies from z = 0.7 to 1.2 with {eBOSS}
  emission-line galaxies},
  \href{https://doi.org/10.3847/1538-4357/aaf9ad}{\emph{The Astrophysical
  Journal} {\bfseries 871} (2019) 147}.

\bibitem{Gonzalez_Perez_2017}
V.~Gonzalez-Perez, J.~Comparat, P.~Norberg, C.M.~Baugh, S.~Contreras, C.~Lacey
  et~al., \emph{The host dark matter haloes of [o ii] emitters at
  0.5 < z < 1.5},
  \href{https://doi.org/10.1093/mnras/stx2807}{\emph{Monthly Notices of the
  Royal Astronomical Society} {\bfseries 474} (2017) 4024–4038}.

\bibitem{Yip2019}
J.H.T.~Yip, X.~Zhang, Y.~Wang, W.~Zhang, Y.~Sun, G.~Contardo et~al., \emph{From
  dark matter to galaxies with convolutional neural networks},  2019.

\bibitem{kasmanoff2020dm2gal}
N.~Kasmanoff, F.~Villaescusa-Navarro, J.~Tinker and S.~Ho, \emph{dm2gal:
  Mapping dark matter to galaxies with neural networks},  2020.

\bibitem{villaescusanavarro2020neural}
F.~Villaescusa-Navarro, B.D.~Wandelt, D.~Anglés-Alcázar, S.~Genel,
  J.M.Z.~Mantilla, S.~Ho et~al., \emph{Neural networks as optimal estimators to
  marginalize over baryonic effects},  2020.

\bibitem{He2019}
S.~He, Y.~Li, Y.~Feng, S.~Ho, S.~Ravanbakhsh, W.~Chen et~al., \emph{{Learning
  to predict the cosmological structure formation}},
  \href{https://doi.org/10.1073/pnas.1821458116}{\emph{Proceedings of the
  National Academy of Sciences of the United States of America} {\bfseries 116}
  (2019) 13825} [\href{https://arxiv.org/abs/1811.06533}{{\ttfamily
  1811.06533}}].

\bibitem{deoliveira2020fast}
R.A.~de~Oliveira, Y.~Li, F.~Villaescusa-Navarro, S.~Ho and D.N.~Spergel,
  \emph{Fast and accurate non-linear predictions of universes with deep
  learning},  2020.

\bibitem{Kodi_Ramanah_2020}
D.~Kodi~Ramanah, T.~Charnock, F.~Villaescusa-Navarro and B.D.~Wandelt,
  \emph{Super-resolution emulator of cosmological simulations using deep
  physical models}, \href{https://doi.org/10.1093/mnras/staa1428}{\emph{Monthly
  Notices of the Royal Astronomical Society} {\bfseries 495} (2020)
  4227–4236}.

\bibitem{TinkerWechsler}
R.H.~Wechsler and J.L.~Tinker, \emph{The connection between galaxies and their
  dark matter halos},
  \href{https://doi.org/10.1146/annurev-astro-081817-051756}{\emph{Annual
  Review of Astronomy and Astrophysics} {\bfseries 56} (2018) 435–487}.

\bibitem{YinLi2021}
Y.~Li, Y.~Ni, R.A.C.~Croft, T.~Di~Matteo, S.~Bird and Y.~Feng,
  \emph{Ai-assisted superresolution cosmological simulations},
  \href{https://doi.org/10.1073/pnas.2022038118}{\emph{Proceedings of the
  National Academy of Sciences} {\bfseries 118} (2021) }
  [\href{https://arxiv.org/abs/https://www.pnas.org/content/118/19/e2022038118.full.pdf}{{\ttfamily
  https://www.pnas.org/content/118/19/e2022038118.full.pdf}}].

\bibitem{white1996}
S.D.M.~{White}, \emph{{Formation and Evolution of Galaxies}},  in
  \emph{Cosmology and Large Scale Structure}, R.~{Schaeffer}, J.~{Silk},
  M.~{Spiro} and J.~{Zinn-Justin}, eds., p.~349, Jan., 1996.

\bibitem{YinLi2021_2}
Y.~Ni, Y.~Li, P.~Lachance, R.A.C.~Croft, T.D.~Matteo, S.~Bird et~al.,
  \emph{Ai-assisted super-resolution cosmological simulations ii: Halo
  substructures, velocities and higher order statistics},  2021.

\bibitem{Vogelsberger2014b}
M.~Vogelsberger, S.~Genel, V.~Springel, P.~Torrey, D.~Sijacki, D.~Xu et~al.,
  \emph{{Introducing the illustris project: Simulating the coevolution of dark
  and visible matter in the universe}},
  \href{https://doi.org/10.1093/mnras/stu1536}{\emph{Monthly Notices of the
  Royal Astronomical Society} {\bfseries 444} (2014) 1518}
  [\href{https://arxiv.org/abs/1405.2921}{{\ttfamily 1405.2921}}].

\bibitem{Bennett_2013}
C.L.~{Bennett}, D.~{Larson}, J.L.~{Weiland}, N.~{Jarosik}, G.~{Hinshaw},
  N.~{Odegard} et~al., \emph{{Nine-year Wilkinson Microwave Anisotropy Probe
  (WMAP) Observations: Final Maps and Results}},
  \href{https://doi.org/10.1088/0067-0049/208/2/20}{\emph{The Astrophysical
  Journal Supplement} {\bfseries 208} (2013) 20}
  [\href{https://arxiv.org/abs/1212.5225}{{\ttfamily 1212.5225}}].

\bibitem{Vogelsberger2014a}
M.~Vogelsberger, S.~Genel, V.~Springel, P.~Torrey, D.~Sijacki, D.~Xu et~al.,
  \emph{{Properties of galaxies reproduced by a hydrodynamic simulation}},
  \href{https://doi.org/10.1038/nature13316}{\emph{Nature} {\bfseries 509}
  (2014) 177} [\href{https://arxiv.org/abs/1405.1418}{{\ttfamily 1405.1418}}].

\bibitem{Goodfellow2014}
I.J.~Goodfellow, J.~Pouget-Abadie, M.~Mirza, B.~Xu, D.~Warde-Farley, S.~Ozair
  et~al., \emph{{Generative Adversarial Networks}},
  \href{https://doi.org/10.1109/ICCVW.2019.00369}{\emph{Proceedings - 2019
  International Conference on Computer Vision Workshop, ICCVW 2019} (2014)
  3063} [\href{https://arxiv.org/abs/1406.2661}{{\ttfamily 1406.2661}}].

\bibitem{Nash1950}
J.F.~Nash, \emph{Equilibrium points in n-person games},
  \href{https://doi.org/10.1073/pnas.36.1.48}{\emph{Proceedings of the National
  Academy of Sciences} {\bfseries 36} (1950) 48}
  [\href{https://arxiv.org/abs/https://www.pnas.org/content/36/1/48.full.pdf}{{\ttfamily
  https://www.pnas.org/content/36/1/48.full.pdf}}].

\bibitem{mirza2014}
M.~Mirza and S.~Osindero, \emph{Conditional generative adversarial nets},
  2014.

\bibitem{Mescheder2018}
L.~Mescheder, A.~Geiger and S.~Nowozin, \emph{{Which training methods for GANs
  do actually converge?}}, {\emph{35th International Conference on Machine
  Learning, ICML 2018} {\bfseries 8} (2018) 5589}
  [\href{https://arxiv.org/abs/1801.04406}{{\ttfamily 1801.04406}}].

\bibitem{Ronneberger2015}
O.~Ronneberger, P.~Fischer and T.~Brox, \emph{{U-net: Convolutional networks
  for biomedical image segmentation}},
  \href{https://doi.org/10.1007/978-3-319-24574-4_28}{\emph{Lecture Notes in
  Computer Science (including subseries Lecture Notes in Artificial
  Intelligence and Lecture Notes in Bioinformatics)} {\bfseries 9351} (2015)
  234} [\href{https://arxiv.org/abs/1505.04597}{{\ttfamily 1505.04597}}].

\bibitem{Giusarma2019}
E.~{Giusarma}, M.~{Reyes Hurtado}, F.~{Villaescusa-Navarro}, S.~{He}, S.~{Ho}
  and C.~{Hahn}, \emph{{Learning neutrino effects in Cosmology with
  Convolutional Neural Networks}}, {\emph{arXiv e-prints} (2019)
  arXiv:1910.04255} [\href{https://arxiv.org/abs/1910.04255}{{\ttfamily
  1910.04255}}].

\bibitem{chen2020learning}
C.~Chen, Y.~Li, F.~Villaescusa-Navarro, S.~Ho and A.~Pullen, \emph{Learning the
  evolution of the universe in n-body simulations},  2020.

\bibitem{Shelhamer2017}
E.~Shelhamer, J.~Long and T.~Darrell, \emph{{Fully Convolutional Networks for
  Semantic Segmentation}},
  \href{https://doi.org/10.1109/TPAMI.2016.2572683}{\emph{IEEE Transactions on
  Pattern Analysis and Machine Intelligence} {\bfseries 39} (2017) 640}
  [\href{https://arxiv.org/abs/1411.4038}{{\ttfamily 1411.4038}}].

\bibitem{odena2016}
A.~Odena, V.~Dumoulin and C.~Olah, \emph{Deconvolution and checkerboard
  artifacts}, \href{https://doi.org/10.23915/distill.00003}{\emph{Distill}
  (2016) }.

\bibitem{adamw2019}
I.~Loshchilov and F.~Hutter, \emph{Decoupled weight decay regularization},
  2019.

\bibitem{ttur2017}
M.~Heusel, H.~Ramsauer, T.~Unterthiner, B.~Nessler and S.~Hochreiter,
  \emph{Gans trained by a two time-scale update rule converge to a local nash
  equilibrium},  2018.

\bibitem{movandenboschwhite2010}
H.~Mo, F.~van~den Bosch and S.~White, \emph{Probing the cosmic density field},
  in \emph{Galaxy Formation and Evolution}, pp.~262--318, Cambridge University
  Press (2010), \href{https://doi.org/10.1017/CBO9780511807244.007}{DOI}.

\bibitem{Jing_2005}
Y.P.~Jing, \emph{Correcting for the alias effect when measuring the power
  spectrum using a fast fourier transform},
  \href{https://doi.org/10.1086/427087}{\emph{The Astrophysical Journal}
  {\bfseries 620} (2005) 559–563}.

\bibitem{nbodykit}
N.~Hand, Y.~Feng, F.~Beutler, Y.~Li, C.~Modi, U.~Seljak et~al., \emph{nbodykit:
  An open-source, massively parallel toolkit for large-scale structure},
  \href{https://doi.org/10.3847/1538-3881/aadae0}{\emph{The Astronomical
  Journal} {\bfseries 156} (2018) 160}.

\bibitem{Springel_2017}
V.~Springel, R.~Pakmor, A.~Pillepich, R.~Weinberger, D.~Nelson, L.~Hernquist
  et~al., \emph{First results from the illustristng simulations: matter and
  galaxy clustering},
  \href{https://doi.org/10.1093/mnras/stx3304}{\emph{Monthly Notices of the
  Royal Astronomical Society} {\bfseries 475} (2017) 676–698}.

\bibitem{doogesh19}
D.~Kodi~Ramanah, T.~Charnock and G.~Lavaux, \emph{Painting halos from cosmic
  density fields of dark matter with physically motivated neural networks},
  \href{https://doi.org/10.1103/physrevd.100.043515}{\emph{Physical Review D}
  {\bfseries 100} (2019) }.

\end{thebibliography}\endgroup
\bibliographystyle{JHEP}

\end{document}